\documentclass[prl,twocolumn,showpacs]{revtex4}
\usepackage{graphicx,natbib}
\newcommand{\vect}[1]{\mathbf{#1}}
\newcommand{\matr}[1]{\mathbf{#1}}

\begin{document}
\title{Fluctuations and redundancy in optimal transport networks}
\author{Francis Corson}
\affiliation{Laboratoire de Physique Statistique, Ecole Normale Sup\'erieure,
UPMC Paris 06, Université\'e Paris Diderot, CNRS,
24 rue Lhomond, 75005 Paris, France.
}
\date{\today}

\begin{abstract}

The structure of networks that provide optimal transport properties
has been investigated in a variety of contexts.
While many different formulations of this problem have been considered,
it is recurrently found that optimal networks are trees.
It is shown here that this result is contingent on the assumption of a stationary flow
through the network.
When time variations or fluctuations are allowed for,
a different class of optimal structures is found,
which share the hierarchical organization of trees yet contain loops.
The transitions between different network topologies
as the parameters of the problem vary are examined.
These results may have strong implications for the structure and formation of natural networks,
as is illustrated by the example of leaf venation networks.
\end{abstract}


\maketitle

The structure of networks that provide optimal transport efficiency
is a question of long-standing interest~\cite{villani-03,stevens-74,ball-98},
with applications ranging from the design of infrastructures such as water and power supply networks 
to the analysis of natural networks such as
the vascular system of plants~\cite{mcculloh-03} and animals~\cite{murray-26}
or river basins~\cite{rodriguez-iturbe-97}.
Optimal networks may be analyzed in terms of their topology,
and, if they are large enough, in terms of their statistical properties.
In this respect, the self-similar structure observed in many natural networks
has strongly stimulated the general investigation of optimal networks.
For natural networks,
the discussion of their functional properties
is also intimately related with that of the patterning mechanisms by which they arise.

While different formulations of the optimal transport problem have been developed in different contexts,
they recurrently yield tree-like structures,
with a single path connecting any two points~\cite{banavar-00,durand-07,bohn-07,bernot-09}
\footnote{We exclude from our analysis the transport of distinct elements
between pairs of locations in the network,
as in telecommunications or land transportation networks.
In that case, a tree-like structure is ineffective
and optimal networks are highly interconnected.}. 
Many natural networks conform with this predicted tree-like organization.
However, it stands in strong contrast to the structure of leaf venation networks,
which contain many loops~\cite{nelson-97}.
This structure could be interpreted as a compromise between
transport efficiency and other requirements,
such as tolerance to damage.
On the other hand, it has been suggested that the redundancy of leaf venation networks
could be an adaptation to the varying water demands of different parts of the leaf~\cite{roth-nebelsick-01}.
In this Letter,
we examine the structure of optimal networks
carrying a flow that varies with time or fluctuates.
Beyond the particular example that motivated this study,
this is a question of general relevance,
since the stationary flow usually assumed in the study of optimal networks is always an idealization.
We show that fluctuations can give rise to a different class of optimal structures,
which contain loops while retaining the hierarchical organization of trees.
Introducing a suitable measure of network redundancy,
we characterize the transitions between different
network topologies as the parameters of the problem are varied.
We discuss the implications of these findings for the structure
and formation of natural networks,
showing how they can account for the presence of loops
in some cases and their absence in others.

When considering networks that transport fluid or electrical current,
a natural formulation of the optimization problem
consists in  minimizing the total dissipation rate
with a limited amount of resources~\cite{durand-07,bohn-07}.
Consider a network formed of vertices $k$ and edges $(k,l)$
having conductances $\kappa_{kl}$.
For simplicity, we consider a regular network
(all edges have length one).
Sources $i_k$ are connected to the vertices, with $\sum_k i_k=0$,
and the edges carry currents $I_{kl}$,
which by Kirchhoff's current law satisfy $i_k=\sum_l I_{kl}$.
The problem consists in minimizing the dissipation
\begin{equation}
J=\sum_{k,l}\frac{I_{kl}^2}{\kappa_{kl}}
\end{equation}
under the constraint that
\begin{equation}
\sum_{k,l}\kappa_{kl}^\gamma=K^\gamma.
\end{equation}
The constant $K^\gamma$ can be interpreted as the amount of material available to build the network,
with the exponent $\gamma$ depending on the nature of the network.
For instance, $\gamma=1$ for electrical wires or porous pipes.
In his analysis of vascular networks,
\citet{murray-26} considered Poiseuille flow in hollow pipes and a fixed volume of fluid,
which corresponds to $\gamma=1/2$.
$\gamma<1$ means that for a given cost,
the dissipation can be lowered by concentrating the flow along links of large conductance.
The case $\gamma>1$ may be considered of little practical relevance,
as it is then more economical to build several parallel links
having a small conductance than a larger one of equivalent capacity.
When the flow through the network is stationary,
it is found that for $\gamma<1$, all local minima of the dissipation
are spanning trees,
while for $\gamma>1$, there is a single, global minimum,
and all edges have non-zero conductance~\cite{banavar-00,bohn-07}.

To study the influence of fluctuations,
we now consider the state of the sources $\vect{i}=(i_1,\ldots,i_N)$
as a random variable with probability density $p(\vect{i})$.
The dissipation in state $\vect{i}$ is
\begin{equation}
J(\vect{i})=\sum_{k,l}\frac{I_{kl}^2(\vect{i})}{\kappa_{kl}},
\end{equation}
where the $I_{kl}(\vect{i})$ are the currents in that state.
The problem is now to minimize the average dissipation
\begin{equation}
<J>=\int J(\vect{i})dp(\vect{i}).
\end{equation}
Introducing a Lagrange multiplier $\lambda$,
the solution is obtained by minimizing
\begin{equation}
\Xi=<J>-\lambda\sum_{k,l}\kappa_{kl}^\gamma.
\end{equation}
In many cases, the currents derive from a potential,
such as pressure or electrical potential,
and are uniquely determined by the sources and conductances.
However, this requirement need not be explicitly stated in the problem.
Indeed, following Thomson's principle~\cite{doyle-84},
when arbitrary currents satisfying Kirchhoff's current law are considered,
the currents that minimize the dissipation derive from a potential.
In all cases, the currents and conductances may thus
be varied independently in the optimization.
Minimizing $\Xi$ with respect to $\kappa_{kl}$ yields
\begin{equation}
\kappa_{kl}=\frac{<I_{kl}^2>^\frac{1}{1+\gamma}}
{\left(\sum_{m,n}<I_{mn}^2>^\frac{\gamma}{1+\gamma}\right)^\frac{1}{\gamma}}K.
\label{kappa}
\end{equation}
This is our first important result,
which relates the conductance of each link in an optimal network
to the mean square current that it carries,
generalizing a similar relation derived for stationary flows~\cite{bohn-07}.

The above relation also serves as the basis for our numerical investigation
of the structure of optimal networks.
Indeed, local minima of the dissipation can be computed by iterating Eq.~\ref{kappa}~\cite{bohn-07}.
For a given set of conductances,
the $<I_{kl}^2>$ can be determined from the source correlations.
The potentials $u_k(\vect{i})$ are related to $\vect{i}$ by
\begin{equation}
i_k=\sum_l \kappa_{kl}\left(u_k(\vect{i})-u_l(\vect{i})\right),
\end{equation}
or in matrix notation $\vect{i}=\matr{C}\vect{u}(\vect{i})$,
and
\begin{equation}
<u_k u_l>=R_{km}R_{ln}<i_m i_n>,
\end{equation}
where $\matr{R}=\matr{C}^{-1}$.
Finally,
\begin{equation}
<I_{kl}^2>=\kappa_{kl}^2\left(<u_k^2>+<u_l^2>-2<u_ku_l>\right).
\end{equation}
The above minimization procedure is started with random conductances.

\begin{figure*}
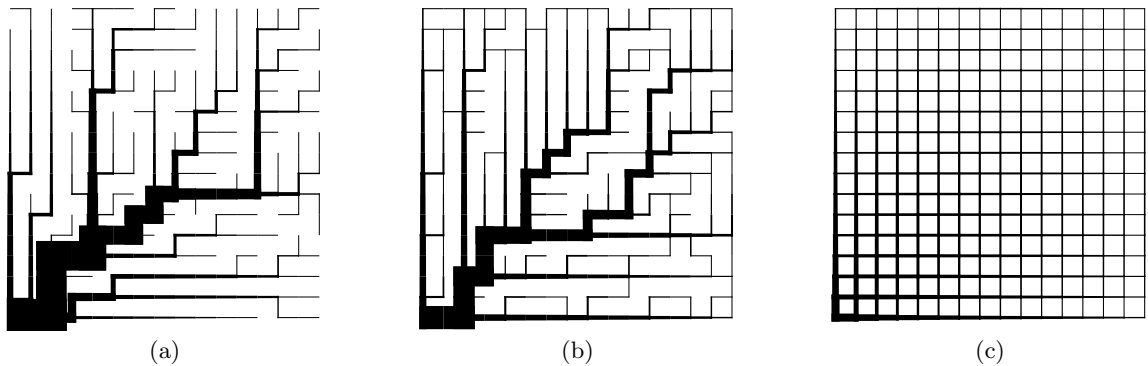

\parbox{.3\textwidth}{
\includegraphics[width=1.75in]{fig1a.eps}\\
(a)}
\parbox{.3\textwidth}{
\includegraphics[width=1.75in]{fig1b.eps}\\
(b)}
\parbox{.3\textwidth}{
\includegraphics[width=1.75in]{fig1c.eps}\\
(c)}
\caption{
Network structures obtained with fluctuating sources for different values of
the parameter $\gamma$ (system size $n=16$, fluctuation amplitude $\sigma=1$).
In each network, the sink is at the lower left corner.
The width of each edge is proportional to the square root of its conductance
(corresponding to the diameter of electrical wires or porous pipes).
(a) Tree-like network ($\gamma=.25$).
(b) Hierarchical network with loops ($\gamma=.75$).
(c) Network with many loops and no hierarchical organization ($\gamma=1.25$;
notice that the network is not perfectly uniform, conductances being higher near the sink).
}
\label{networks}
\end{figure*}

In what follows, an $n\times n$ square network ($N=n^2$),
with $N-1$ sources ($2\le k\le N$) and a sink ($k=1$) in one corner is considered.
The sources have unit average and uncorrelated fluctuations of amplitude $\sigma$:
$<i_k>=1$ and $<i_k i_l>=1+\sigma^2\delta_{kl}$, $2\le k,l\le N$,
where $\delta$ denotes the Kronecker delta function
(the $<i_k i_1>$ are obtained from $\sum_k i_k=0$).
For sources that are randomly switched on and off with equal probability,
$\sigma=1$.
The network topologies observed with constant sources,
trees and rather uniform networks with many loops,
are recovered for $\gamma$ small enough (Fig.~\ref{networks}(a))
and for $\gamma>1$ (Fig.~\ref{networks}(c)),
respectively.
Our central result is that in an intermediate range of $\gamma$, however,
hierarchical networks with loops are obtained (Fig.~\ref{networks}(b)).
An intuitive interpretation of this result is that
$\gamma<1$ favors grouping the flow along edges of high conductance,
while loops provide alternate routes to accommodate fluctuations.
The local minima obtained for $\gamma<1$ depend on the initial conditions.
For small systems, however, we have checked that local minima containing loops
yield a lower dissipation than the optimal tree
by constructing all spanning trees rooted at the sink~\cite{gabow-78}.
Note also that local minima are relevant where self-organized processes are concerned,
as in the case of natural networks.

\begin{figure}
\setlength{\unitlength}{1in}
\begin{picture}(0,0)
\put(.4,1.25){(a)}
\end{picture}
\includegraphics{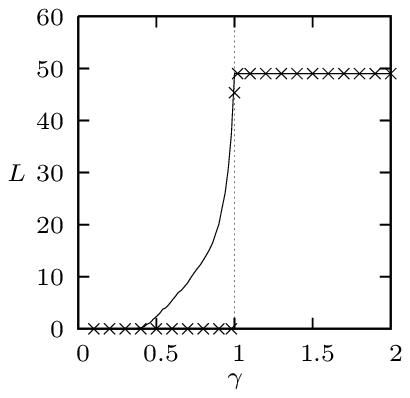}
\begin{picture}(0,0)
\put(.4,1.25){(b)}
\end{picture}
\includegraphics{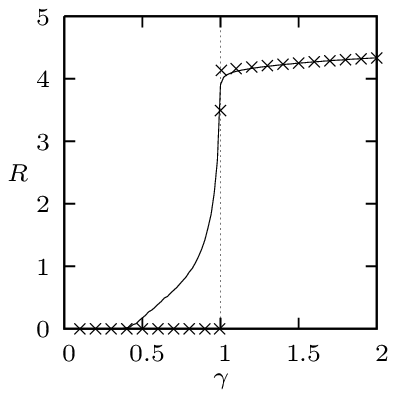}
\caption{
Evolution of network structure vs. $\gamma$
with constant (crosses) and fluctuating (lines) sources
($n=8$, $\sigma=1$, averages over $100$ realizations).
(a) Number of loops $L$.
For $\gamma>1$, all conductances are non-zero and $L=(n-1)^2$.
(b) Redundancy $R$. Note that $R$ is not constant for $\gamma>1$.
}
\label{transition}
\end{figure}

To investigate the transitions between these different network topologies,
it is suitable to introduce a measure of network redundancy.
A simple such measure is the number of loops $L$,
which can be determined using Euler's formula,
$L=1+\textrm{number of edges}-\textrm{number of vertices}$,
where only edges that exceed a threshold conductance are counted.
Fig.~\ref{transition}(a) shows that loops appear above a finite threshold $\gamma_c$,
and their number progressively increases to reach its maximum as $\gamma\to 1$.
The number of loops, however, is not an ideal measure of network redundancy,
as it is a purely topological one, and depends on an arbitrary threshold.
These limitations can be overcome by introducing an entropy-based measure of redundancy,
which is similar to measures of reliability appearing in the literature on water distribution networks~\cite{ang-05},
and could be of general interest in the analysis of complex networks.
First, the redundancy $r_{ab}$ between two vertices $a$ and $b$ is defined,
using an analogy with random walks.
Consider the currents $I_{kl}$ corresponding to a unit current
injected into the network at $a$ and extracted at $b$
($i_k=\delta_{ka}-\delta_{kb}$)
and transition probabilities
\begin{equation}
P_{kl}=\left\{\begin{array}{ll}
\frac{I_{kl}}{\sum_{I_{km}>0}I_{km}}, &I_{kl}>0,
\\
0, &I_{kl}\le 0.
\end{array}\right.
\end{equation}
The probability of any path \mbox{$\vect{k}=(k_1,\ldots,k_l)$} from $a$ to $b$
($k_1=a$, $k_l=b$) is
\begin{equation}
P(\vect{k})=\prod_{1\le j<l}P_{k_j k_{j+1}},
\end{equation}
and the redundancy $r_{ab}$ is defined as
the entropy of the path distribution:
\begin{equation}
r_{ab}=-\sum_\vect{k} P(\vect{k})\log P(\vect{k}).
\end{equation}
This expression can be transformed into a sum over the nodes,
$r_{ab}=\sum_{i}P_i S_i$,
where $P_i$ is the total inflow into node $i$ and
$S_i=-\sum_j P_{ij}\log P_{ij}$. 
$r_{ab}$ is equal to $0$ if there is a single path from $a$ to $b$,
to $\log n$ if there are $n$ independent paths of equal conductance from $a$ to $b$,
and can take arbitrary positive values in general situations
($e^{r_{ab}}$ can be interpreted as a generalized number of paths
between $a$ and $b$).
In the present case of identical sources and a sink,
a sensible measure of the overall redundancy $R$ of the network
is the average redundancy between any source and the sink,
i.e.
\begin{equation}
R=\frac{1}{N-1}{\sum_{2\le k\le N}r_{kN}}.
\end{equation}

As shown on figure~\ref{transition}(b),
the variation of the redundancy $R$ with the parameter $\gamma$
is qualitatively similar to that of the number or loops.
The redundancy is 0 when there are no loops,
and gradually increases above the threshold $\gamma_c$.
Note however that it is not constant for $\gamma>1$,
and only progressively tends to its maximum
as $\gamma\rightarrow+\infty$ and the conductances become uniform.
The value of the maximum is of order $n$,
reflecting the exponential growth of the number of possible
paths in the network.

\begin{figure}
\includegraphics{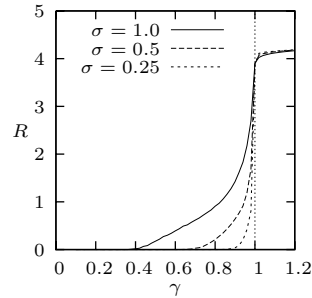}
\caption{
Redundancy $R$ vs. $\gamma$ for different
fluctuation amplitudes ($n=8$).
}
\label{noise}
\end{figure}

\begin{figure}
\setlength{\unitlength}{1in}
\mbox{
\begin{picture}(0,0)
\put(.5,.5){(a)}
\end{picture}
\includegraphics{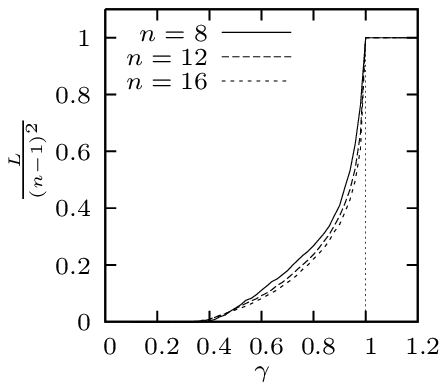}
\begin{picture}(0,0)
\put(.5,.5){(b)}
\end{picture}
\includegraphics{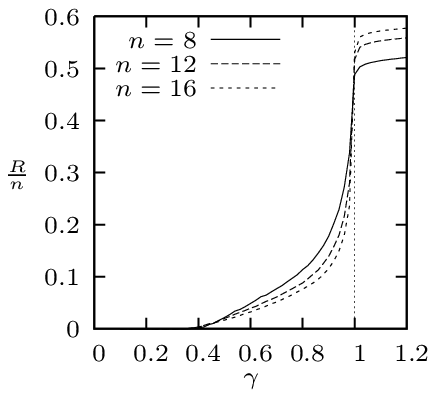}
}
\caption{
Dependence of network structure on system size ($\sigma=1$).
(a) Number of loops $L$ rescaled by $1/(n-1)^2$.
(b) Redundancy $R$ rescaled by $1/n$.
}
\label{size}
\end{figure}

Using this measure of redundancy,
we now turn to an analysis of the transitions
between different network structures as the parameters
of the model are varied.
Fig.~\ref{noise} shows the dependence of the redundancy $R$
on the amplitude of the fluctuations.
Expectedly, the transition is sharper ($\gamma_c$ is larger)
for smaller amplitudes.
Although a limited range is accessible numerically
(each step of the minimization procedure involves the inversion of an $N\times N$ matrix),
we have also examined the dependence of optimal network structure on system size.
Fig.~\ref{size} shows the number of loops and redundancy as a function of $\gamma$
for different system sizes.
We find that $\gamma_c$ appears to remain constant.
The number of loops scales as $(n-1)^2$,
suggesting a constant average loop area.
On the other hand, the normalized redundancy appears
to increase less rapidly above $\gamma_c$
for larger values of $n$.
This may be due to a weaker redundancy on the large scales,
due to the averaging out of the uncorrelated fluctuations.

In summary,
we have shown that fluctuations dramatically change the optimal structure
of transport networks,
favoring the presence of loops.
The optimal structure is determined by the correlations
between the currents injected into the network.
We restricted our analysis to uncorrelated fluctuations,
and further investigations should address the influence of correlations,
which can be expected to increase redundancy on larger scales.
It would also be of interest to examine the statistical properties of the structures obtained.
One the one hand,
optimal trees are generally found to be self-similar~\cite{rodriguez-iturbe-97}.
This property might allow an estimation of the threshold for the appearance of loops
by balancing the cost of inserting a new link and the corresponding dissipation reduction. 
On the other hand, one could ask whether self-similar structures
can also be obtained in reticulate networks.
This might be the case when the fluctuations
exhibit power-law correlations.
In another line,
it has been shown for stationary flows
that the geometry of network junctions satisfies special relations
when dissipation is minimized with respect to the locations of the vertices~\cite{durand-06},
and one could attempt to generalize these relations in the presence of fluctuations.

An application of our results may be found in the structure of natural networks,
and in particular in the reticulate structure of leaf venation networks.
Indeed, the rate of transpiration in the different regions of a leaf
is known to fluctuate over time~\cite{peak-04}.
In addition, leaf veins are bundles of smaller vessels,
with transport properties similar to a porous medium,
corresponding to a value of $\gamma$ close to 1
(yet smaller than 1 since
the vessel diameters increase with vein diameter).
This is the range where the structure of optimal networks
is most sensitive to fluctuations.
Interestingly, long-range correlations have been observed
in the fluctuations of leaf transpiration rate~\cite{messinger-07},
which could explain why leaf venation networks exhibit redundancy at every scale.
On the other hand, the vascular networks of animals
can be described by the smaller value $\gamma=1/2$,
and a tree-like structure would be predicted
even if moderate fluctuations are taken into account.
Our results may also be relevant to the formation 
of reticulate networks such as leaf venation networks.
Indeed, Eq.~\ref{kappa} shows
that optimal structures may be obtained through local adaptation
of the conductance to the flow.
Returning to a functional perspective,
another benefit of redundancy is tolerance to damage,
and it would be of interest to compare the structures obtained here with
those found when explicitly optimizing for tolerance to damage.

\acknowledgments{
I thank Arezki Boudaoud and Mokhtar Adda-Bedia for helpful discussions
and suggestions on the manuscript, and Pradeed Kumar
for comments on the manuscript.
This work was supported by the MechPlant project of
European Union's New and Emerging Science and Technology program.
}

\bibliography{refs}
\end{document}